\documentclass[prl,aps,floats,showpacs,
                preprint
                ]{revtex4}

\usepackage{graphicx}

\begin{document}

\title{Surface Oscillations in Overdense Plasmas Irradiated by
Ultrashort Laser Pulses}
\author{A. Macchi$^1$}\email{macchi@df.unipi.it}
\author{F. Cornolti$^1$}
\author{F. Pegoraro$^1$}
\author{T. V. Liseikina$^1$}
\author{H. Ruhl$^2$}
\author{V. A. Vshivkov$^3$}
\affiliation{$^1$Dipartimento di Fisica and INFM (sezione A), 
Universit\'{a} di Pisa, Italy}
\affiliation{$^2$Max-Born Institut f\"ur Quantenoptik, Berlin, Germany}
\affiliation{$^3$Institute of Computational Technologies of SD-RAS,
Novosibirsk, Russia}

\begin{abstract}
The generation of electron surface oscillations
in overdense plasmas irradiated at normal
incidence by an intense laser pulse is
investigated. Two-dimensional ($2$D) particle-in-cell
simulations
show a transition from a planar, electrostatic oscillation
at $2\omega$, with $\omega$
the laser frequency, to a $2$D electromagnetic oscillation
at frequency $\omega$ and wavevector
$k>\omega/c$. 
A new electron parametric instability, 
involving the decay of a 1D electrostatic
oscillation into two surface waves, 
is introduced to explain
the basic features of the $2$D oscillations.
This effect leads to the rippling of the
plasma surface within a few laser cycles, 
and is likely to have a strong impact on laser
interaction with solid targets.
\end{abstract}

\maketitle

The interaction of sub-picosecond, high-intensity laser pulses
with solid targets is of great relevance to the generation
of bright sources of energetic radiation 
as well as a test bed for Fast Ignitor physics. 
Since for a solid target the electron density
$n_e \gg n_c$, where
$n_c=1.1 \times 10^{21} \text{ cm}^{-3}/[\lambda/\mu\text{m}]^2$
is the cut-off density for laser
propagation at the wavelength $\lambda$,
the laser-plasma coupling occurs at the target surface
over a narrow region with a depth of the order of the
skin length $d_p \ll \lambda$. The laser force on the plasma 
has both secular components (leading to plasma acceleration, profile
steepening and hole boring) and oscillating components;  these 
latter drive
an oscillatory motion of the ``critical'' 
surface where $n_e=n_c$, that acts as a ``moving mirror'' 
leading to the appearance of high harmonics in the reflected light
\cite{bulanov}.

Experiments \cite{feurer}
and simulations \cite{wilks,ruhl1,macchi}
suggest that either pre-imposed or self-generated deformations
of the target surface strongly affect laser energy absorption.
Evidence for small-scale
deformations comes from the
wide spreading of the reflected radiation
observed in experiments 
\cite{norreys,tarasevitch,vonderlinde}
at high intensities ($\approx {10^{18} \text{ W cm}^{-2}}$) and even
for pulse durations as short as $35 \text{ fs}$ \cite{tarasevitch}.
This suggests that surface rippling involves some ``fast''
mechanism related to electron motion rather than Rayleigh-Taylor-like 
(RT) hydrodynamic instabilities driven by the strong
target acceleration and occuring on time scales of the ion motion.
Seeding of density ``ripples'' by electron instabilities
in the underdense plasma region in front 
of the target was evidenced in simulations 
\cite{wilks} and 
investigated theoretically in
\cite{cadjan}.

In this Letter, we show with numerical simulations that
electron surface oscillations (ESOs)
may grow for a step density profile (no underdense region
present) much faster than the typical
time scale of ion motion,
leading to an oscillatory
``rippling'' of the critical surface.
To interpret the simulations we present a model for
a new parametric instability, based on the ``decay''
of pumped one-dimensional electrostatic oscillations into
two electron surface waves.
This represents a new nonlinear mechanism of surface wave excitation by
laser pulses different from previously investigated models 
\cite{dragila}.
Its potentially strong impact on the disruption
of ``moving mirrors'' for high harmonics generation (HHG) and the
production of fast electron jets are discussed.

We use two-dimensional ($2$D)
particle-in-cell (PIC) simulations to study the dynamics of
the ESOs with proper spatial and temporal resolution.
In particular, in order to evaluate the
frequency of the ESOs,
the complete output of 2D fields was produced eight times for 
each laser cycle. 
In the simulations reported, the laser pulse is
normally incident, impinges from the left
on the $x$-axis, and has
a wavelength $\lambda=0.25 \mu\text{m}$
($n_c=1.6 \times 10^{22} \text{ cm}^{-3}$),
a uniform spatial profile in the transverse
($y$) direction
and a temporal profile that rises  for three cycles and then
remains constant. The laser pulse is ``$s$-polarized'', i.e.
its electric field is in the $z$ direction normal
to the simulation plane.
The plasma has immobile ions, 
initial temperature $T_e=5 \mbox{ keV}$
and a step-like density profile
filling the
right part of the box ($x>0$)
A numerical box $12 \lambda \times 4 \lambda$ is
taken with a spatial resolution equal to the 
Debye length 
$\lambda_D=\sqrt{T_e/4\pi n_o e^2}$,
where $n_o$ is the initial density, 
and 25 particles per cell are used.
We will focus on two typical simulations (run~1 and run~2).
Introducing the dimensionless irradiance
$a_o=0.85(I\lambda^2/10^{18} \text{ W}\mu\text{m}^2/\text{cm}^2)^{1/2}$
in run~1 we take $a_o=1.7$ and $n_o=5n_c$, while in run~2 we take
$a_o=0.85$ and $n_o=3n_c$.

The contours of the electron density $n_e(x,y)$  in Fig.~\ref{fig1}
for run~1 at times $t=8,10,12,$ and $14$ laser cycles
from the run start, show the evolution of the surface dynamics over
several laser cycles.
Correspondingly, the  space-time contours of
$n_e(x,y=y_i,t)$ at $y_1=2.0\lambda$ and $y_2=1.875\lambda$
in Fig.~\ref{fig2}  show the temporal
behavior of the surface oscillations. Initially, the surface
oscillation is planar, i.e. uniform along $y$,
and has a frequency $2\omega$,
being $\omega=2\pi c/\lambda$ the laser frequency.
It is natural to identify this 1D motion as the
``moving mirror'' driven by the longitudinal
${\bf j} \times {\bf B}$ force at $2\omega$.
In the compression phase, electrons pile up in a
narrow layer where the peak density is $n_e\simeq 2n_o$;
the electrostatic field $E_x^{(2\omega)}$ (not shown)
is positive and counteracts the
${\bf j} \times {\bf B}$ force.
In the expansion phase
electrons are dragged out into vacuum, forming a
``cloud'' of negative charge with a
negative electrostatic field.

The growth of surface ``ripples''
can be  observed in Fig.\ref{fig1}. At $t=10$,
they have small wavelengths ($\simeq 0.1 \lambda$),
while at  $t \geq 12$ they evolve into
a steady oscillation with  wavelength
$\lambda_s \simeq 0.5 \lambda$ (Fig.\ref{fig1}) and
frequency $\simeq \omega$, which
is superimposed to the oscillation at $2\omega$ (Fig.~\ref{fig2}).
This ``period doubling'' effect is also observed in run~2, for which
the density contours at $t=16.5$, $17$, $17.5$, and $18.0$ 
are shown in Fig.~\ref{fig3}: it is evident that
the curvature of the density layer
is inverted each half laser cycle.
In run~2 the oscillation amplitude and the density compression
is lower than in run~1, and the deformation wavelength
is larger ($\lambda_s \simeq 0.75 \lambda$).
In the following we will discuss only the long-wavelength
structures oscillating at $\omega$ and refer to them
as ($2$D)ESOs. The $2$DESOs
are ``standing'',
i.e.  not  propagating  along $y$.
From Fig.\ref{fig2} we see that while at
$y_1$ the amplitude is close to its maximum, at
$y_2 = y_1-\lambda_s/4$ there is no evident growth
of the oscillation at $\omega$, while a
weakening of the
$2\omega$ oscillation is observed.
Results (not shown) for $p$-polarization 
at normal incidence (i.e, with the electric field
of the laser pulse in the $y$ direction) 
are qualitatively similar, the main difference
being that the plasma ``plumes'' extending into 
vacuum are bent by the laser field in the plane.

To our knowledge, there are no previous simulations or models
describing surface ``ripples'' of the electron density 
oscillating at the laser frequency for normal incidence.
In the ``moving mirror''
motion a superposition of $\omega$ and $2\omega$ motions occurs
only for oblique incidence and $p$-polarization,
because in such a
case both the electric and magnetic forces
have components normal to the
surface \cite{vonderlinde2}. In ref.\cite{plaja} grating-like
surface inhomogeneities oscillating at $2\omega$ and 
induced by the magnetic force at oblique incidence 
were studied.
Stationary surface ripples in the ion density 
were observed 
in simulations \cite{cadjan}
at much longer times \footnote{In principle
the long time evolution of 2DESOs may lead
to ion density ripples similar to those
observed in \cite{cadjan}, which however
have wavelength $\lambda_s > \lambda$, while the 
opposite inequality holds in our model of 2DESOs based on surface wave
excitation.}
and interpreted on the basis of
a hydrodynamic model 
for an underdense, homogeneous, neutral 
plasma which leads (for $s$-polarization 
only) to an instability
pumped by the secular part of the ponderomotive force.
Our explanation is that the
$2$DESOs are generated
by a parametric decay of the forced 1D oscillation
with frequency $2\omega$ and transverse wavevector $k_y=0$
into two electron surface waves (ESWs)
$(\omega_1,k_1)$ and $(\omega_2,k_2)$.
Using the matching conditions 
$k_1+k_2=0$ and $\omega_1+\omega_2=2\omega$
we find that
the two overlapping ESWs thus form a standing oscillation with frequency
$\omega=\omega_1=\omega_2$ and wavevector $k=k_1=-k_2$.

A cold fluid, non-relativistic 2D model
of the parametric excitation of ESWs
has been developed.
Here we report only a brief description of
this model while a detailed derivation will
be published elsewhere
\cite{macchi2}. Firstly,
the quiver motion along $z$ is solved and  a system of
2D Maxwell-Euler
equations is obtained where the laser action enters via the 
ponderomotive force.
The electron fluid density and velocity are written as
$n_e=n_o(x)+\epsilon \delta
n_e^{(2\omega)}(x,t) + \epsilon^2 \delta
n_e^{(\omega)}(x,y,t) $  and ${\bf v}_e=\epsilon
V_x^{(2\omega)}(x,t)\hat{\bf x} +\epsilon^2 {\bf v}^{(\omega)}(x,y,t)$,
where  $\epsilon \sim a_o^2 (n_c/n_e)$
is a small expansion parameter. The terms at  $2\omega$ describe the
electrostatic,
1D ``moving mirror'' oscillation, that acts as a pump for the
instability,
and is assumed to be unperturbed by the ESWs. The terms at
$\omega$ are the superposition of
two ESWs:
\begin{equation}
{\bf v}^{(\omega)}=e^{-i\omega t}
\left(\tilde{\bf v}_{+k}e^{iky}+
   \tilde{\bf v}_{-k}e^{-iky}\right )/2 +\text{ c.c.}
\,  ,
\end{equation}
where $\tilde{\bf v}_{\pm k}=\tilde{\bf v}_{\pm k}(x,t)$
varies slowly in time.
To order $\epsilon^2$, the coupling between 1D and 2D modes may
be neglected, so that one obtains the usual dispersion relation
for ``H''  surface waves
propagating along a density discontinuity:
\begin{equation}
k^2 c^2=\omega^2~
(\omega_p^2-\omega^2)/(\omega_p^2-2\omega^2),
\label{eq:SW}
\end{equation}
where
$\omega_{p}^2=4\pi n_o e^2/m_e$
is the plasma frequency.
The evanescence length of the ESWs in the plasma is
$L_{SW} =(c/\omega)(1-2\omega^2/\omega_p^2)^{1/4}
(1-\omega^2/\omega_p^2)^{-1/2}$.
Notice that for electron surface waves 
$\nabla\cdot{\bf E}^{(\omega)}=-4\pi e\delta n_e^{(\omega)}(x,y,t)=0$,
and that their maximum frequency is $\omega_{max}=\omega_p/\sqrt{2}$, 
so that the matching 
conditions may be satisfied only if $\omega<2\omega_{max}$, i.e.
$n_e>2n_c$.

Inserting the value of
the laser frequency in (\ref{eq:SW})  we find that the expected 
wavelength of deformations in
run~2 is  $\lambda_{s}=2\pi/k \simeq 0.71 \lambda$, in good agreement
with the simulation.
For run~1 one finds
$\lambda_s \simeq 0.87 \lambda$, quite larger than the
numerical result. This is not surprising since our expansion
procedure is not applicable for the parameters of run~1, where
the interaction is in the relativistic regime.
This may cause, for instance, a lowering of
the effective plasma frequency by the 
(time-averaged) relativistic factor
$\gamma_o \simeq \sqrt{1+a_o^2/2}$, mostly due to the
relativistic quiver motion along $z$.
By  replacing $\omega_p^2$ by
$\omega_p^2/\gamma_0$ in (\ref{eq:SW}),
for run~1 we obtain $\lambda_s \simeq 0.55 \lambda$, much closer
to the numerical result. 

By keeping only terms up order $\epsilon^3$ in
the Euler equation and neglecting feedback effects on
the 1D motion, the Euler equation for the ESW velocity
is $\partial_t {\bf v}^{(\omega)}=- e {\bf E}^{(\omega)}/m_e
+\epsilon {\bf a}^{(\omega)}_{NL}$ 
where ${\bf a}^{(\omega)}_{NL}$ describes the nonlinear coupling
with the 1D motion: 
\begin{equation}
{\bf a}^{(\omega)}_{NL}=
-V_x^{(2\omega)} \partial_x {\bf v}^{(\omega)} 
                   -v_x^{(\omega)}\partial_x V_x^{(2\omega)}
                    \hat{\bf x}
            +\frac{eV_x^{(2\omega)}}{m_e c}
B_z^{(\omega)}\hat{\bf y}.
\end{equation}
Using this equation and Poynting's theorem
the rate of growth of the surface energy 
for the $2$DESOs was evaluated as
$\Gamma \equiv U^{-1} \partial_tU$,
where $U$ is the energy density per wavelength 
of the two ESWs and the cycle-averaged variation is
\begin{eqnarray}
&\partial_t U=
\int\!dx
\left\langle {\bf v}^{(\omega)}
    \cdot \left(e\delta \tilde{n}_e^{(2\omega)}
              \tilde{\bf E}^{(\omega)}
              +m_e n_o\partial_t \tilde{\bf v}^{(\omega)}\right)\right\rangle
\nonumber \\
&=\frac{1}{4}\int_{0}^{+\infty}\!dx 
\tilde{\bf v}^{*}_{+k}
  \cdot \left (e\delta \tilde{n_e}^{(2\omega)}\tilde{\bf E}^{*}_{-k}
  \right.
-m_e n_o \tilde{v}^{*}_{x,-k}\partial_x V_x^{(2\omega)}
  \nonumber \\
&-m_e n_o \tilde{V}_x^{(2\omega)}\partial_x \tilde{\bf v}^{*}_{-k}
+n_o \frac{e}{c} \tilde{V}_x^{(2\omega)}\tilde{B}^{*}_{z,-k}\hat{\bf y}
\left. \right )+ \text{ c.c.}      
\end{eqnarray}
Substituting in the integrand
for the expressions of the (unperturbed) ESW fields,
finally one obtains the growth rate as
\begin{equation}
\Gamma \simeq
{4\omega a_o^2}
\frac{(\alpha-1)^{3/2}}{\alpha|\alpha-4|[(\alpha-1)^2+1](\alpha-2)^{1/2}}
\label{eq:rate}
\end{equation}
where $\alpha=n_e/n_c=\omega_p^2/\omega^2$.
The denominator $(\alpha-4)$
is actually due to  the resonant excitation of longitudinal plasmons at
$\omega_p=2\omega$, which makes $V_x^{(2\omega)}$ very large
and invalidates our ordering assumptions near resonance.
We note that $\Gamma$ diverges also for $\alpha \rightarrow 2$;
however, in this limit the ESW wavelength is very small and thus 
one expects a strong damping by thermal effects neglected in the 
cold fluid model.

Our model is valid at normal incidence
for both $s$- and $p$-polarizations (in the sense of the 2D 
simulations), 
since even if the laser electric 
field is parallel to the $k$-vectors of the ESWs
there is no resonant coupling of the quiver motion with the ESWs.
The transverse oscillations are however observed more clearly
for $s$- rather than for $p$-polarization, since in the latter case
the motions in the laser and the ESW fields overlap in the 
polarization plane.
For oblique incidence and $p$-polarization, in addition to the $2\omega$
motion the laser drives the moving mirror at frequency $\omega$, 
which may lead to the pumping of two ESW sidebands around the 
frequency $\omega/2$ \cite{macchi3}.
This latter case 
is directly relevant to experiments on HHG,
where the surface rippling 
observed at high intensities 
may pose a limit to HHG as an efficient XUV source. 
The normal incidence
results  already indicate their  potential relevance to the
interpretation  of this phenomenon.
First, the growth rate is maximum
in conditions ``optimal'' for HHG,
i.e. when the moving mirror oscillation is driven at 
high velocities, which requires a moderate density ``shelf''
rather than solid densities. In these conditions,
at the time of maximum pulse intensity the
plasma has a finite gradient at the surface $n_e=n_c$ which is in
turn steepened by the radiation pressure.
Therefore, the plasma profile may look
rather similar to that assumed in our simulations. 
In such conditions
the ESO instability can be much faster than RT instabilities. 
In fact,
even for accelerations of order  $g \simeq 10^{20} \text{ cm/s}^2$
as measured in this regime \cite{ACC},
the typical RT growth rate 
$\Gamma_{RT} \simeq \sqrt{k_{RT}g} \simeq (140 \text{ fs})^{-1}$
for $2\pi/k_{RT} \simeq \lambda/2= 0.125 \mu\text{m}$, is
much slower than the rise of the  $2$DESOs which occurs over a few
$\text{ fs}$ in the simulations.

Finally, the results of run~1 give an indication on 
the scaling of the growth rate in the relativistic 
regime, not accessible to our analytical model; the growth rate 
may increase strongly for relativistic intensities
due to the decrease of the effective plasma frequency,
producing a stronger rippling of the surface, as consistent 
with the simulation results and experiments that suggest that
the instability threshold is close to $a_o = 1$.

The ESOs also have a substantial impact on fast
electron generation. Fig.~\ref{fig4}
shows phase space distributions at times $t=8.5$ and $13.5$ laser cycles
for run~1.
At early times, the momentum distribution is  uniform in
$y$, with no accelerated particles in $p_y$ and
most energetic electrons having $p_x \simeq 2m_ec$.
At later times, when the $2$DESOs have grown, stronger forward
acceleration occurs near the maxima of the oscillation, showing
that most oscillatory energy has been transferred to
the unstable $2$D modes. Correspondingly, strong acceleration
in $p_y$ also occurs. The momentum distribution
in $p_x$ is very regular, with  only a minority of electrons
``outrunning'' the oscillation. This suggests that the
generation of fast electrons is correlated
with the nonlinear evolution and ``breaking'' of the density
oscillations, when the amplitude of the latter exceeds the
screening length. This may give a spatial ``imprint'' 
on the transverse structure
of the fast electron currents. 
In previous simulations one may clearly 
observe a spatial correlation between electron jets
and ``corrugations'' at the surface 
\cite{lasinski,sentoku}.
When penetrating into the bulk the jets 
may either merge
or drive current filamentation 
instabilities
\cite{sentoku,califano}, thus producing different 
spatial scales.

The PIC simulations were performed at the CINECA supercomputing
facility
(Bologna, Italy), sponsored by the INFM supercomputing initiative.

\begin{figure}
\caption{Contours 
of normalized electron density $n_e/n_c$ 
for run~1 ($a_o=1.7$, $n_o=5n_c$) at
various 
times (see plot labels) in laser cycle units.
Only a small portion of the 
simulation box around
the target surface is 
shown.}
\label{fig1}
\end{figure}

\begin{figure}
\caption{Space-time evolution of $n_e(x,y=y_l,t)$
at $y_1=1.875/\lambda$ (left) and $y_2=2.0\lambda$ (right) for run~1
(see Fig.\ref{fig1}).
The position of the $n_e=n_c$ surface is evidenced by a black
contour line.}
\label{fig2}
\end{figure}

\begin{figure}
\caption{Same as Fig.~\ref{fig1} for run~2 ($a_o=0.85$, $n_o=3n_c$).}
\label{fig3}
\end{figure}

\begin{figure}
\caption{Phase space projections for run~1 in the
$(y,p_x)$ (top),
$(y,p_y)$ (middle), and
$(x,p_x)$ (bottom) planes
for $t=8.5$ (left) and $t=13.5$ (right) laser cycles.}
\label{fig4}
\end{figure}

\end{document}